\documentclass[
reprint,
amsmath,amssymb,superscriptaddress,
aps,
prl
]{revtex4-1}

\usepackage{graphicx}
\usepackage{dcolumn}
\usepackage{bm}
\usepackage{color}
\usepackage{float}
\usepackage{verbatim}
\restylefloat{table}

\usepackage[normalem]{ulem} 

\begin{document}

\title{Composite Fermions with a Warped Fermi Contour}
\date{\today}

\author{M. A.\ Mueed}
\author{D.\ Kamburov}
\author{Yang\ Liu}
\author{M.\ Shayegan}
\author{L. N.\ Pfeiffer}
\author{K. W.\ West}
\author{K. W.\ Baldwin}
\affiliation{ Department of Electrical Engineering, Princeton University, Princeton, New Jersey 08544, USA}
\author{R.\ Winkler}
\affiliation{Department of Physics, Northern Illinois University, DeKalb, Illinois 60115, USA}
\affiliation{Materials Science Division, Argonne National Laboratory, Argonne, Illinois 60439, USA}

\begin{abstract}
Via measurements of commensurability features near Landau filling factor $\nu=1/2$, we probe the shape of the Fermi contour for hole-flux composite fermions confined to a wide GaAs quantum well. The data reveal that the composite fermions are strongly influenced by the characteristics of the Landau level in which they are formed. In particular, their Fermi contour is \textit{warped} when their Landau level originates from a hole band with significant warping.
\end{abstract}

\maketitle

At very low temperatures, clean two-dimensional (2D) carrier systems manifest signatures of many-body interaction under a strong perpendicular magnetic field ($B_{\perp}$). One such phenomenon is the fractional quantum Hall effect (FQHE) which is elegantly explained in the framework of composite fermions (CFs), exotic quasi-particles composed of charged particles bound to an even number of magnetic flux quanta \cite{Jain.2007, Jain.PRL.1989, Halperin.PRB.1993}. This flux attachment cancels the external magnetic field at Landau levels (LL) filling factor $\nu=1/2$, causing CFs to behave as if they are at $B_{\perp}=0$ and occupy a Fermi sea with a well-defined Fermi contour \cite{Jain.2007, Jain.PRL.1989, Halperin.PRB.1993, Willett.PRL.1993, Kang.PRL.1993, Goldman.PRL.1994, Smet.PRL.1996, Kamburov.PRL.2012, Kamburov.PRL.2013, Kamburov.PRB.2014, Kamburov.PRL.2014}. Away from $\nu=1/2$, CFs feel the effective magnetic field $B^*_{\perp}=B_{\perp}-B_{\perp,1/2}$, where $B_{\perp,1/2}$ is the field at $\nu=1/2$ \cite{Footnote1}.

The role of anisotropy in FQHE has been featured in many recent studies \cite{Balagurov.PRB.2000, Gokmen.NatPhys.2010, Bo.PRB.2012, Kamburov.PRL.2013, Papic.PRB.2013, Kun.PRB.2013, Kamburov.PRB.2014}. In particular, it has been a puzzle whether CFs inherit energy band anisotropy from the carriers at low magnetic field. In the simplest scenario, the band properties of the low-field particles should not map onto CFs because the latter are primarily a product of interaction. However, measurements on AlAs quantum wells (QWs) containing 2D electrons with an elliptical Fermi contour and anisotropic transport revealed that CFs also exhibit \textit{resistance} anisotropy \cite{Gokmen.NatPhys.2010}. Although this suggests a possible inheritance of energy band properties by CFs, anisotropic resistance could also be caused by anisotropic scattering even if the Fermi contour for CFs were isotropic. Without a direct measurement of CFs' Fermi contour, it remains unclear if there is any transference of anisotropic band dispersion from electrons. In this Letter, we address this question through direct measurements of the \textit{Fermi contour} for hole-flux CFs confined to wide GaAs QWs where the (zero-field) 2D hole Fermi contour is significantly warped. We find that the warping is qualitatively transferred to CFs. Our additional data, taken with an applied parallel magnetic field ($B_{||}$), provide evidence that the warping depends on the LL in which the CFs are formed.

\begin{figure}
\includegraphics[width=.48\textwidth]{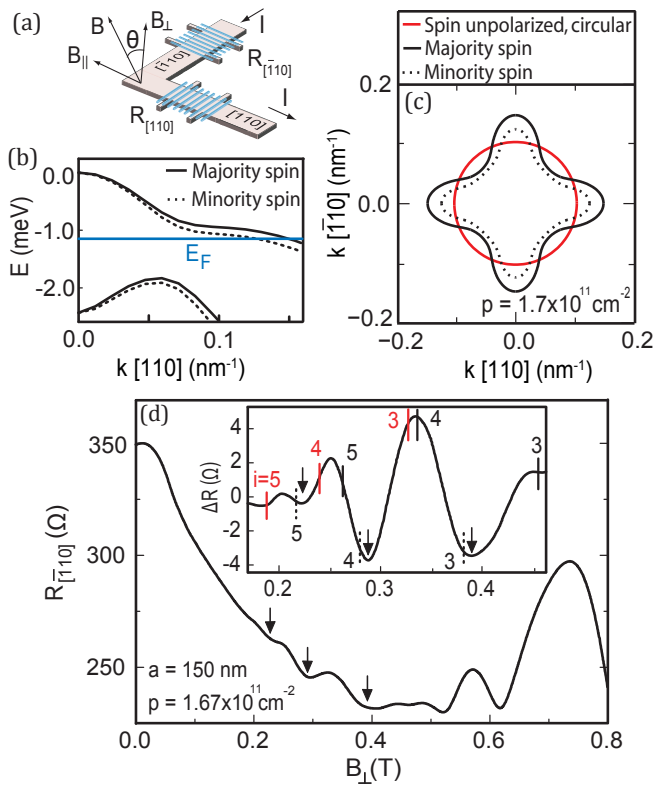}
\caption{\label{fig:Fig1} (color online) (a) Sample
schematics. The electron-beam resist grating covering the top surface of each Hall bar arm is shown as blue stripes. (b) Dispersions for the two lowest, spin-split subbands along [110] for a wide GaAs QW, calculated self-consistently for $B= 0$. (c) Calculated Fermi contours, exhibiting significant warping. (d) Magnetoresistance trace from the $[\overline{1}10]$ arm of a sample with period $a=150$ nm. The trace for the [110] arm (not shown) is similar. Inset: Pronounced COs are seen after subtracting the background resistance using a second-order polynomial fit. Observed resistance minima do not match the $B_\perp$-positions expected for a spin-degenerate, circular, Fermi contour (red tick-marks) but lie in between the black solid and dotted tick-marks which denote the expected COs positions for the majority and minority spin Fermi contours, respectively. All expected positions are calculated according to $2R_{c}/a=i-1/4$ (see text).}
\end{figure}

We studied 2D hole systems (2DHSs) confined to a 35-nm-wide symmetric GaAs QW, grown by molecular beam epitaxy on a (001) GaAs substrate. The QW, located 131 nm below the surface, is flanked on each side by 95-nm-thick Al$_{0.24}$Ga$_{0.76}$As spacer layers and C $\delta$-doped layers. The 2DHS density ($p$) is $\simeq 1.67\times10^{11}$ cm$^{-2}$, and its mobility is $\simeq 10^{6}$ cm$^{2}$/Vs. As shown in Fig. 1(a), we fabricated a Hall bar with two perpendicular arms oriented along [110] and $[\overline{1}10]$. The arms are covered with periodic stripes of negative electron-beam resist which, through the piezoelectric effect in GaAs, produce a density modulation of the same period in the 2DHS \cite{Kamburov1.PRB.2012, Kamburov2.PRB.2012, Skuras.APL.1997, Endo.PRB.2000, Endo.PRB.2001, Kamburov.PRL.2012, Kamburov.PRB.2014, Kamburov.PRL.2013, Kamburov.PRL.2014}. We measured, at $T=0.3$ K, the longitudinal resistances along the two arms in purely perpendicular and also in tilted magnetic fields, with $\theta$ denoting the angle between the field direction and the normal to the 2D plane (Fig. 1(a)). The sample was tilted around $[\overline{1}10]$ so that $B_{||}$ was always along [110].

Figures 1(b) and (c) show the energy band dispersions and the Fermi contours of a 2DHS confined to a wide GaAs QW \cite{footnote4, Liu.PRB.2014}, based on an $8\times8$ Kane Hamiltonian \cite{Winkler} which combines the Dresselhaus spin-orbit coupling and the non-parabolicity of the 2D hole bands. As seen in Fig. 1(c), the Fermi contour is significantly warped as a result of severe mixing between the heavy-hole (HH) and light-hole (LH) states \cite{Broido.PRB.1985}. Spin-orbit coupling also causes the contours of two different spin species to split. As a result of warping, the Fermi wave vectors ($k_{F}$) for both majority and minority spin contours along [110] and $[\overline{1}10]$ are larger than $k_{F}$ of a circular Fermi contour which contains the same number of (spin-unpolarized) 2D holes.

We use commensurability oscillations (COs) to probe the Fermi contour shapes of both holes and hole-flux CFs. The COs are manifested in the magnetoresistance as a minimum whenever the quasi-classical cyclotron orbit diameter $2R_{c}$ of the particles becomes commensurate with the period of the density modulation, $a$. Since $2R_{c}=2{\hbar}k_{F}/eB_\perp$, the $B_\perp$-positions of COs resistance minima provide a direct measure of $k_F$. For a spin-unpolarized, circular Fermi contour, the expected positions of these minima are given by the electrostatic commensurability condition, $2R_{c}/a=i-1/4$ \cite{Weiss.EurophysL.1989, Winkler.PRL.1989, Gerhardts.PRL.1989, Beenakker.PRL.1989, Beton.PRB.1990, Peeters.PRB.1992, Mirlin.PRB.1998}, where, $i$ is an integer and $k_{F,cir}={\sqrt {2{\pi}p}}$. In Fig. 1(d) we show the low-field magnetoresistance of our 2DHS along $[\overline{1}10]$ which allows us to deduce $k_{F}$ along [110] \cite{Footnote3}. The red tick-marks in the inset denote the expected $i=3,4,5$ $B_\perp$-positions for $k_{F, cir}$ while the solid and dotted black tick-marks are for majority and minority spin Fermi contours (see Fig. 1(c)), respectively. Clearly, the positions of observed resistance minima (arrows) in Fig. 1(d) inset do not match the red marks. Each minimum, however, is close to the dotted black tick mark of a given $i$, suggesting that the Fermi contour is warped. This observation also implies that the Fermi contour agrees better with the minority spin contour, consistent with previous studies on other 2DHSs \cite{Kamburov2.PRB.2012}. With this interpretation, we deduce a value of $\simeq20\%$ for the observed warping of the Fermi contour. (We define the warping as the ratio of $k_{F}$ along [110] or $[\overline{1}10]$ over $k_{F, cir}$.) We note that, generally, warping is significantly more pronounced in wide wells such as those studied here than in narrower QWs studied previously \cite{Kamburov2.PRB.2012}.

Having established a significant warping in our 2DHS Fermi contours, we now turn to the Fermi contour of $\nu=1/2$ CFs. As seen in the magnetoresistance data of Fig. 2, there are two pronounced minima on the sides of $\nu=1/2$, flanked by shoulders of rapidly increasing resistance. These two minima correspond to the commensurability of CFs' cyclotron orbit diameter $2R_{c}^{*}$ with $a$. Quantitatively, for a circular CF Fermi contour, the positions of these resistance minima are given by the $magnetic$ commensurability condition, $2R_{c}^{*}/a=5/4$ \cite{Willett.PRL.1999, Smet.PRB.1997, Smet.PRL.1998, Mirlin.PRL.1998, Oppen.PRL.1998, Smet.PRL.1999, Zwerschke.PRL.1999, Kamburov.PRL.2013, Kamburov.PRL.2014, Kamburov.PRL.2012, Kamburov.PRB.2014}, where $2R_{c}^{*}={2\hbar}k_{F, cir}^{*}/eB_{\perp}^{*}$ is the quasi-classical cyclotron orbit of CFs at the effective magnetic field $B_{\perp}^{*}$, $k_{F,cir}^{*}={\sqrt{4{\pi}p^{*}}}$, and $p^{*}$ is the CF density. The expression for $k_F^*$ assumes full spin-polarization at high fields of $\simeq$ 14 T. Recent studies have established that, in the vicinity of $\nu=1/2$, $p^{*}$ is equal to the $minority$ carrier density, namely $p^{*}=p$ for $B_{\perp}^{*}>0$ and $p^{*}=[(1-\nu)/\nu]p$ for $B_{\perp}^{*}<0$ \cite{Kamburov.PRL.2014}. In Fig. 2 inset, we mark the expected field positions (red tick-marks) of CF commensurability minima for a circular Fermi contour based on the minority density in the lowest LL. The positions of the observed resistance minima (vertical arrows) are measurably farther from $B_{\perp,1/2}$ than the red marks, providing clear evidence that CFs have a warped Fermi contour. Based on Fig. 2 data, and also similar data taken on three other samples, we deduce a warping ($k^{*}_{F}/k_{F,cir}^{*}$) of $\sim 15\%$ for the CFs. This is comparable to, but somewhat smaller than the warping we measure for the hole Fermi contour (Fig. 1(d)), suggesting that CFs inherit some warping in their Fermi contour from the LL in which they are formed.

\begin{figure}
\includegraphics[width=.44\textwidth]{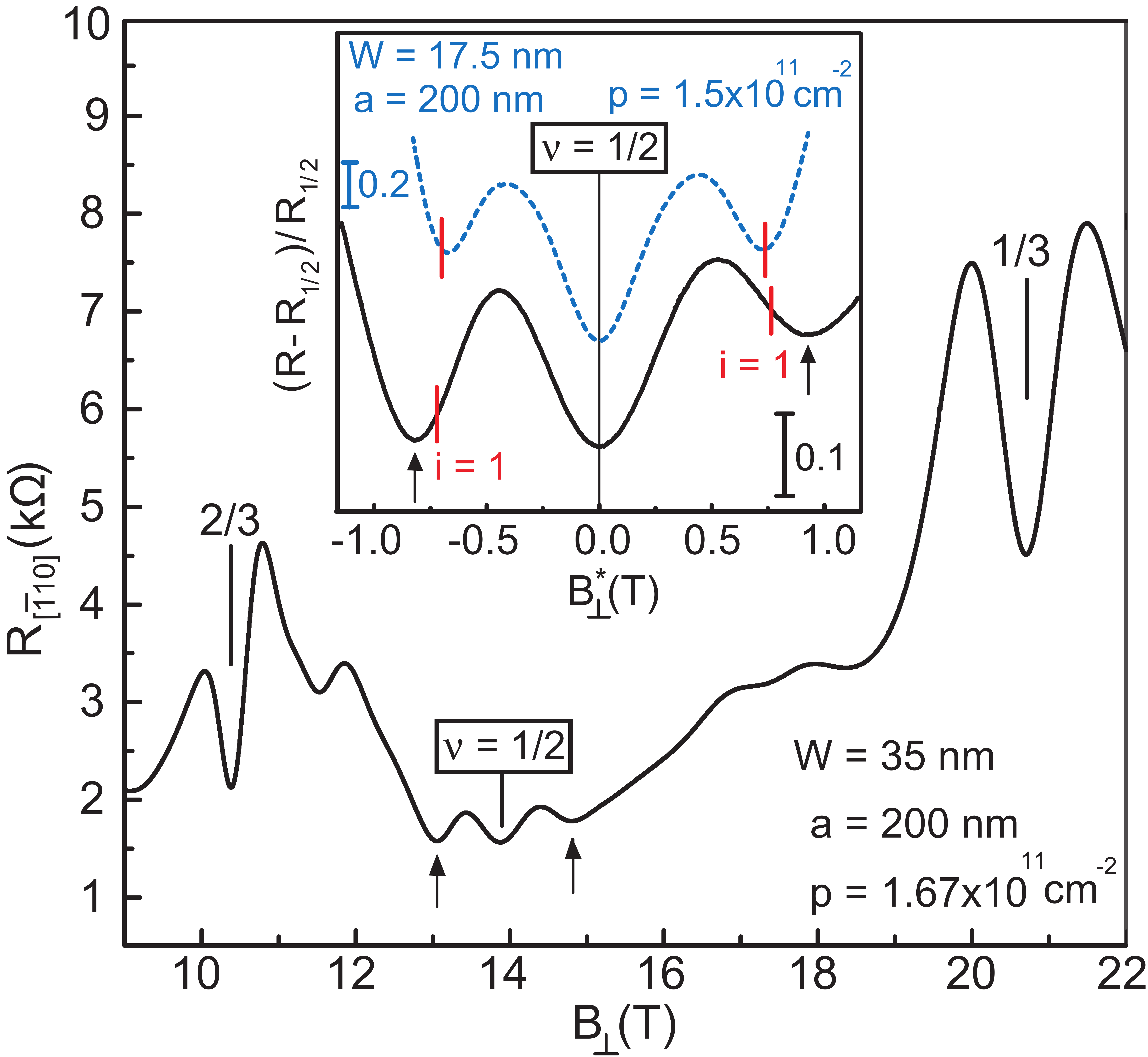}
\caption{\label{fig:Fig2} (color online) Magnetoresistance trace from the $[\overline{1}10]$ Hall bar for a GaAs QW of width $W$ = 35 nm. The two prominent minima near $\nu=1/2$ are signatures of commensurability of CF cyclotron orbit diameter with the period ($a=200$ nm) of the density modulation. Inset: Enlarged trace near $\nu=1/2$ shows that the positions of the minima are measurably farther from $B_{\perp}^{*}=0$ than expected for a circular Fermi contour of fully spin-polarized CFs (marked by vertical red tick-marks). For comparison, we also include data (dashed blue trace) from a narrower 2DHS ($W$ = 17.5 nm) \cite{Kamburov.PRL.2012, Kamburov.PRL.2013, Kamburov.PRL.2014}. In this case, similar to their 2D hole counterparts near zero magnetic field, CFs also show no warping in their Fermi contour, as illustrated by commensurability minima near $\nu=1/2$ which agree well with the red tick-marks that are based on a circular Fermi contour.}
\end{figure}

We investigate the Fermi contour warping of CFs further by utilizing the crossing of the two lowest-energy LLs at large $B_{\perp}$ \cite{Fischer.PRB.2007, Graninger.PRL.2011, Liu.PRB.2014}. Such a crossing, prevalent in wide QWs where the energy separation between HH and LH subbands is small, can be tuned by either changing the 2DHS density or, at a fixed density, by applying a parallel magnetic field $B_{||}$ \cite{Liu.PRB.2014, Graninger.PRL.2011}. Here we present data, taken as a function of $B_{||}$, demonstrating how the character of the LL in which CFs are formed influences the CFs' Fermi contour warping. In Figs. 3(a) and (b) we summarize the evolution of the magnetoresistance features near $\nu=1/2$ as a function of $\theta$. There are pronounced CF commensurability features consistent with a warped Fermi contour at $\theta=0^{\circ}$, along both $[110]$ and $[\overline{1}10]$. As $\theta$ increases to $\sim26^{\circ}$, the resistance near $\nu=1/2$ increases by a factor of $\simeq 2$, and the magnetoresistance traces become monotonic, losing all commensurability features. For $\theta>30^{\circ}$, however, the resistance near $\nu=1/2$ again becomes comparable to that of the $\theta=0^{\circ}$ trace and, remarkably, the commensurability features around $\nu=1/2$ reappear and become very pronounced.

\begin{figure}
\includegraphics[width=.48\textwidth]{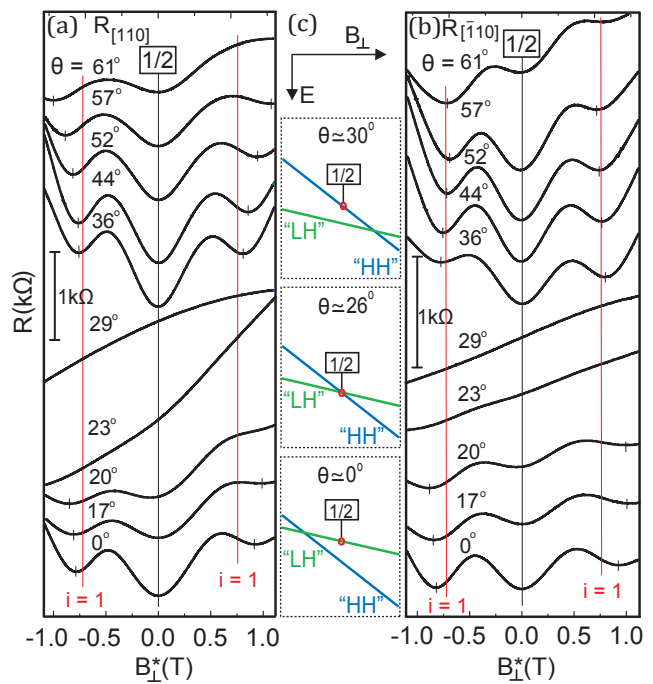}
\caption{\label{fig:Fig3} (color online) (a),(b) Evolution of the
magnetoresistance in the vicinity of $\nu = 1/2$ measured along [110] and $[\overline{1}10]$, respectively. Traces are shifted vertically for clarity and tilt angle $\theta$ is given for each trace. The vertical red lines mark the expected positions of the CF commensurability resistance minima if the CF cyclotron orbit were circular. (c) Crossing between the LLs with LH and HH character as a function of $\theta$. Note that the lowest LL, in which $\nu = 1/2$ CFs are formed, changes from "LH" to "HH" as $\theta$ increases.}
\end{figure}

\begin{figure}
\includegraphics[width=.48\textwidth]{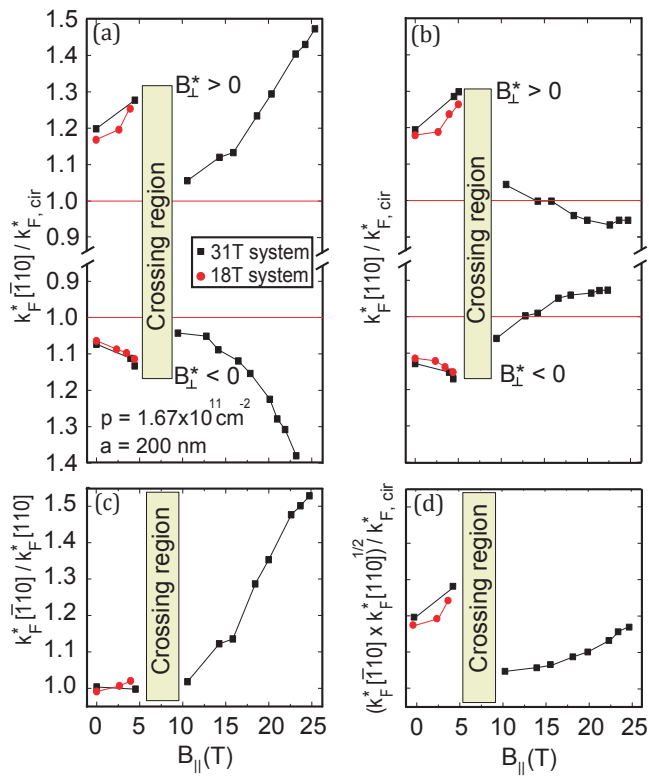}
\caption{\label{fig:Fig4} (color online) (a),(b) Measured values
of the CF Fermi wave vectors $k_F^{*}$ along $[\overline{1}10]$ and [110], normalized to $k^*_{F,cir}$, as a function of $B_{||}$ for both positive and negative $B^*_{\perp}$. Filled squares (black) and filled circles (red) are from measurements in two different systems. (c) Relative anisotropy of the CF Fermi contour is plotted as the ratio of $k_F^{*}$ along $[\overline{1}10]$ and [110]; data for $B^*_{\perp}>0$ were used. (d) The geometric mean of the measured $k_F^{*}$ along $[\overline{1}10]$ and [110], divided by $k^*_{F,cir}$, as a measure of how much the Fermi contour deviates from an ellipse. The yellow region signifies the LL crossing.}
\end{figure}

To explain Fig. 3 data, we focus on the nature of the two lowest-energy LLs. Although these LLs originate from states which are HH-like at the subband edge $k=0$, their exact characteristic is complex due to the admixture of LH and HH states at finite $k$. For simplicity we will refer to these LLs as ``LH" and ``HH", respectively. Figure 3(c) shows a qualitative picture for the crossing between ``LH" and ``HH", which can be explained by the different in-plane (cyclotron) effective masses of ``LH" and ``HH": ``HH" which has a smaller in-plane effective mass compared to ``LH" increases in energy more rapidly with $B_{\perp}$ than ``LH", leading to the crossing at sufficiently large $B_{\perp}$ \cite{Liu.PRB.2014}. At $\theta=0^{\circ}$, $\nu=1/2$ starts out to the right of the crossing and the CFs are formed in ``LH" (Fig. 3(c) lower panel). As $\theta$ increases, the confining potential due to $B_{||}$ becomes stronger thus lowering ``LH" in energy with respect to ``HH". As a result, the crossing position moves closer to $\nu=1/2$ (Fig. 3(c) middle panel) \cite{Liu.PRB.2014}. The increase in resistance near $\nu=1/2$, when $\theta \sim26^{\circ}$ indeed comes about because, when the crossing occurs very close to $\nu=1/2$, the ground-state at $\nu=1/2$ becomes insulating; this is best seen in data taken at lower temperatures on another 2DHS confined to a 35-nm-wide GaAs QW \cite{Liu.PRB.2014}. When the sample is tilted to higher $\theta$ so that the crossing moves well to the right of $\nu=1/2$ (Fig. 3(c) top panel), the resistance near $\nu=1/2$ decreases by a factor of $\simeq 2$, suggesting that the insulating phase has passed and the CFs now form in ``HH".

According to the above discussion, the character of the LL in which CFs are formed changes from ``LH" to ``HH" in the course of the crossing \cite{Footnote6}. This change could affect the CFs' Fermi contour warping. Using the relation, ${2\hbar}k_{F}^{*}/eB_{\perp}^{*}=5/4$, we extract the size of $k_F^*$ along $[\overline{1}10]$ and [110] from the positions of the CF commensurability minima along [110] and $[\overline{1}10]$, respectively. The deduced values of $k_F^*$, normalized to $k^*_{F, cir}$ and plotted as a function of $B_{||}$ in Figs. 4(a) and 4(b), provide a measure of the CF Fermi contour warping. For $B_{||}=0$, $k^*_{F}/k^*_{F, cir}>1$ is consistent with warping of the CF Fermi contour. With increasing $B_{||}$, $k^*_{F}$ increases along both directions until the LL crossing region sets in at $B_{||}\simeq5$ T (see Figs. 4 (a) and 4(b)). Once the commensurability features reappear past the crossing region ($B_{||}\gtrsim10$ T), $k^*_{F}/k^*_{F, cir}$ clearly shows a smaller value than at $B_{||}\simeq5$ T. This drop in $k_F^*$ coincides with the Fermi level at $\nu=1/2$ having moved from the ``LH" to the ``HH" LL.

After the crossing, for $B_{||}>10$ T, $k_F^*$ increases along $[\overline{1}10]$, but decreases along [110] as a function of $B_{||}$, implying an elongation of the CF Fermi contour along $[\overline{1}10]$. This elongation, summarized in Fig. 4(c) plot, results from the coupling between the out-of-plane (orbital) motion of CFs and $B_{||}$, qualitatively confirming previous findings \cite{Kamburov.PRL.2013, Kamburov.PRB.2014}. However, unlike in previous studies, when we plot the geometric mean of $k_F^{*}[\overline{1}10]$ and $k_F^{*}[110]$, normalized to $k^*_{F, cir}$, we find significant deviations from unity (Fig. 4(d)), implying that the Fermi contour is not elliptical and is severely warped. Figure 4(d) also shows that the warping is more severe just to the left of the crossing region compared to the right. This observation suggests a more severe warping when the CFs are formed in ``LH" than in ``HH". It is tempting to attribute this to more severe warping in the ``LH" LL but, unfortunately, there are no theoretical calculations that determine the precise nature of hole LLs in a tilted magnetic field. While our observations indicate that CFs are not decoupled from the underlying crystal structure and inherit band properties such as warping, a quantitative description awaits future theoretical calculations.

We acknowledge support through the DOE BES (DEFG02-00-ER45841) for measurements, and the Gordon and Betty Moore Foundation (Grant No. GBMF4420), Keck Foundation, and the NSF (DMR-1305691, ECSS-1001719 and MRSEC DMR-1420541) for sample fabrication and the NSF (Grant DMR-1310199) for calculations. Work at Argonne was supported by DOE BES (DE-AC02-06CH11357). Our work was partly performed at the National High Magnetic Field Laboratory (NHMFL), which is supported by NSF (DMR-1157490), the State of Florida, and the DOE. We thank S. Hannahs, T. Murphy, J. Park, G. Jones and A. Suslov at NHMFL for valuable technical support and J. K. Jain for illuminating discussions.


\begin{thebibliography}{99}
\bibitem{Footnote1}Throughout this paper, we use `` * " to denote
the CF parameters.
\bibitem{Footnote6}We caution again that the strong mixing of the hole states, especially in the presence of both perpendicular and parallel fields, precludes us from assigning pure HH and LH character to the crossing LLs.
\end{thebibliography}
\end{document}